\documentclass{article}
\begin{document}

\title{
Power Law and Entropy
}

\author{
Sachio Hirokawa \& Eisuke Ito
}
\date{April 22, 2015}


\maketitle



Shannon\cite{Shannon1951} and Yavuz\cite{Yavuz1974} estimated the entropy of
real documents. This note drives an upper bound of entropy from the power law.

\vspace{0.5cm}
Let $D$ be a set of documents and $fr(w)$ be the number of occurrences of a word $w$
in the document. Given an interger $k$, we denote by $F(k)$
the number of words whose frequency is $k$, i.e.,
$F(k) = \sharp \{ w \in D \mid fr(w) = k\}$.
The entropy of the document set $D$
is defined by $E(D) = \Sigma_{w \in D} p(w) log(p(w))$, where
$p(w) = fr(w)/N$ and $N = \Sigma_{w \in D} fr(w)$.

\vspace{0.5cm}
\noindent {\bf Theorem}
If $F(k)$ follows the power law: $log(F(k)) = -a*log(k) + b$ with $a \ge 1$,
then $E(D) \le e^{b(1-\frac{1}{a})}(\frac{b}{a}+1)$.

\vspace{0.5cm}
\noindent {\bf Proof} Since $a \ge 1$, we have
$-log(k)+\frac{b}{a} \le log(F(k)) \le -log(k)+b$. Thus we have,
$\frac{e^{\frac{b}{a}}}{k} \le F(k) \le \frac{e^b}{k}$
and $e^{\frac{b}{a}} \le F(k)*k \le e^b$. Let $M$ be the maximal word frequency.
The the sum of word frequencies $N$ can be evalated as follows.
$N = \Sigma_{w} fr(w) = \Sigma_{k=1}^{M} \Sigma_{fr(w)=k} fr(w)$
$= \Sigma_{k=1}^{M} \Sigma_{fr(w)=k} k$ 
$= \Sigma_{k=1}^{M} F(k)*k \ge \Sigma_{k=1}^{M} e^{\frac{b}{a}}$
$= M e^{\frac{b}{a}}$
Therefore, we have $N \ge M e^{\frac{b}{a}}$ and $\frac{M}{N} \le e^{-\frac{b}{a}}$.
Now, we can evaluate the entropy as follows:
$E(D)$ $=$ $-\Sigma_{w} p(w) log(p(w))$
$=$ $-\Sigma_{w} \frac{fr(w)}{N} log(\frac{fr(w)}{N})$
$=$ $-\Sigma_{k=1}^{M} \Sigma_{fr(w)=k} \frac{k}{N} log(\frac{k}{N})$
$=$ $-\Sigma_{k=1}^{M} F(k)*\frac{k}{N} log(\frac{k}{N})$
$\le$ $-\Sigma_{k=1}^{M} e^b \frac{1}{N} log(\frac{k}{N})$
$\le$ $-e^b \int_{0}^{\frac{M}{N}} log(x) dx$
$\le$ $-e^b \int_{0}^{e^{-\frac{b}{a}}} log(x) dx$
$\le$ $-e^b \left[  x(log(x)-1)+C \right]_0^{e^{-\frac{b}{a}}}$
$=$ $e^{b(1-\frac{1}{a})}(\frac{b}{a}+1)$


\end{document}